\mag=\magstephalf
%\mag=\magstep1
\pageno=1
\input amstex
%\baselineskip = 1.0 true cm
\documentstyle{amsppt}
\TagsOnRight
\interlinepenalty=1000
\hsize=6.4truein
\voffset=23pt
\vsize =9.8truein
\advance\vsize by -\voffset
%\baselineskip = 1.0 true cm
\nologo

\NoBlackBoxes
\font\twobf=cmbx12
\define \ii{\roman i}
\define \ee{\roman e}
\define \dd{\roman d}

%\define \qcl{{\roman{qcl}}}
\define \qcl{{\roman{}}}
\define \reg{{\roman{reg}}}

\define \SS{{\Cal S}}
\define \CC{{\Cal C}}
\define \tp{{t^+}}
\define \tm{{t^-}}
\define \tpm{{t^\pm}}
\define \DD{/\!\!\!\!\Cal{D}}
\define \tvskip{\vskip 0.5 cm}

{\centerline{\bf{On Density of State of Quantized Willmore 
Surface}}}

{\centerline{\bf{---A Way to Quantized Extrinsic String in 
$\Bbb R^3$---}}}

\author
\endauthor
\affil
Shigeki MATSUTANI\\
2-4-11 Sairenji, Niihama, Ehime, 792 Japan \\
\endaffil
\endtopmatter
%\baselineskip= 0.8 true cm

\subheading{Abstract}

Recently I quantized an elastica with Bernoulli-Euler functional
in two-dimensional space using the modified KdV hierarchy.
 In this article, I will quantize a Willmore surface, 
 or equivalently 
a surface with the Polyakov extrinsic curvature
action, using the modified Novikov-Veselov (MNV) equation.
In other words, I show that the density of state of the 
partition function for  the quantized Willmore surface 
is expressed by volume of 
a subspace of the moduli of the MNV equation.

%\subheading{PACS numbers}:
%\baselineskip= 0.8 true cm

\document

%\newpage
%\baselineskip= 0.8 true cm

\tvskip
\centerline{\twobf \S 1. Introduction }
\tvskip

In series of works [1-7], I have considered the correspondence 
between an immersed object
and the Dirac operator confined there. The Dirac operator confined
 in an immersed object is uniquely determined by
the procedure which I proposed [1-4] and can be regarded as
the representation matrix of the symmetry of the immersed object
 [1-7].
I had been studying it mainly on an elastica in a plane [1-6]
 and showed that the Dirac operator confined in an elastica is 
 identified with the Lax operator of the 
modified KdV equation while the mathematical deformation of the 
elastica
obeys the modified KdV hierarchy [6].
By investigating other quantum equations [8-9],
I conjectured that such correspondence between the Dirac operator
 and geometry
can be extended to higher dimensional immersed objects [2,3,4].

Couple years ago Konopelchenko [10,11] discovered that a conformal
 surface $\SS$ immersed in three dimensional flat space $\Bbb R^3$
  obeys the Dirac 
equation, which I will call 
Konopelchenko-Kenmotsu-Weierstrass-Enneper (KKWE) [10-15]
equation here,
$$
    \partial f_1 = V f_2, \quad \bar \partial f_2= -V f_1,
        \tag 1-1
$$
where
$$
       V:= \frac{1}{2}\sqrt{\rho} H, \tag 1-2
$$
$H$ is the mean curvature of the surface $\SS$ parameterized
 by complex $z$ and $\rho$ is  the conformal metric induced from 
 $\Bbb R^3$.
The KKWE equation completely exhibits the immersed geometry as 
the old Weierstrass-Enneper  equation expresses the minimal 
surface [10-15]. In ref.[16] I showed that it is identified with 
the Dirac operator confined in the surface
$\SS$ and by quantizing the Dirac field I found that the quantized
 symmetry of the Dirac operator is also in agreement with the 
 symmetry of the surface itself [17].
In other words,  this KKWE equation is the equation which 
I conjectured before [2,3] and I had been searching for.
 Even though for more general surface, which is not
conformal, the KKWE type equation was discovered by Burgress 
and Jensen [18] following my  prescriptions [1], 
their equation is not easy to deal with and I could not find
 meaningful results.
However the KKWE equation is very useful to investigate the 
immersed object and in terms of  (1-1), Konopelchenko, 
Taimanov and other Russian group found
non-trivial results related to the immersed surface [10-15,19,20].

By physical investigating the KKWE equation and its quantized 
version, the Willmore functional [21,22] and the modified 
Novikov-Veselov (MNV) equation
naturally appears [10-15,17,19,20]. The Willmore functional 
is given as [21,22], 
$$
           W = \int_{\Cal S}  \dd \text{vol}\  H^2 , \tag 1-3
$$
where "$\dd \text{vol}$" is a volume form of the surface $\SS$.
The harmonic map associated with this functional has been studied in
the differential geometry [21,22].

On the other hands, Polyakov introduced an extrinsic curvature
 action in the string theory
and the theory of two-dimensional gravity from renomalizability [23].
However his action is just the Willmore functional (1-3).
Thus his program  recently  was investigated by Carroll and 
Konopelchenko [19]
and Grinevich and Schmidt [20] using KKWE equation (1-1). 
The main theme of this article is to quantize the Willmore surface
but I emphasis that it means the study on the quantization of 
the Polyakov extrinsic curvature action.

It should be noted that the elastica problem of $\Bbb R^2$
 has very similar structure 
of the Willmore surface problem of $\Bbb R^3$ [10-14].
Corresponding to the Willmore functional (2-20), there is 
Bernoulli-Euler functional for an elastica [24],
$$
           E=\int \dd q^1\ k^2, \tag 1-4
$$
where $k$ is a curvature of the elastica [24]. 
While the Willmore surface is related to 
the modified Novikov-Veselov (MNV) equation, 
the elastica is related to the modified KdV equation
[1-7,25-27].

Recently I exactly quantized the elastica of the Bernoulli-Euler
functional (1-4) preserving its local length [25].
Then I found that its moduli is completely represented 
by the MKdV equation and closely related to the two-dimensional 
quantum gravity [28-30].
The quantized elastica obeys the MKdV equation and at a 
critical point, the Painlev\'e equation of the first kind appears
 [25] while in the quantized two-dimensional gravity which is 
 defined at a critical point
of the discrete tiling model, there appears the Painlev\'e equation
 of the first kind
with the KdV hierarchy [28-30].

In this article instead of the local length preserving, 
I will impose that the surface 
preserves its complex structure and 
will  quantize the Willmore functional.
Then I will show that the MNV hierarchy appears  as
the quantized motion of a Willmore surface in the path integral.

The organization of this article is as follows.
Section 2 reviews the argument of 
the quantized elastica following to
that in ref.[25]. In \S 3, I will quantize the Willmore surface
and then the density of states of the Willmore functional
is given as the volume of the MNV equation.
Section 4 gives the discussion for the results.

\tvskip
\centerline{\twobf \S 2. Quantization of  Elastica}
\tvskip

I will denote by $\CC$ a shape of the elastica  embedded 
in a complex plane $\Bbb C $
and by $X(s)$ its affine vector [6]: 
$$
           S^1 \ni s \mapsto X(s) \in \CC \subset \Bbb C  ,
           \quad X(s+L)=X(s),
           \tag 2-1
$$
where $L$ is the length of the elastica.
I will fix  the metric of the curve $\CC$ induced from 
 the natural metric of $\Bbb C $; $\dd s = \sqrt{\dd X \dd \bar X}$.
The Frenet-Serret relations are expressed as [6,25-27]
$$
    \psi_0:=\exp(\ii \phi/2)=\sqrt{\partial_s X},  \tag 2-2 
$$
$$
           \pmatrix \partial_s & v \\
          v & -\partial_s \endpmatrix 
          \pmatrix \psi_0 \\ \ii \psi_0\endpmatrix
          =0 , 
          \quad v:= \frac{1}{2} k :=\frac{1}{2}\partial_s \phi,
                    \tag 2-3
$$
where $\phi$ is a real valued function of $s$ and $k$ is 
the curvature of the curve $\CC$,  
$\phi(s+L)=\phi(s)$ and $k(s+L)=k(s)$.
  
The energy functional of  the elastica, which I will call 
Bernoulli-Euler functional here [24], is given as
$$
       E=\int_0^L \dd s\  k^2=4 \int_0^L \dd s\  v^2 ,\tag 2-4
$$
and  shape of a static elastica is realized as its stationary point
 satisfied with the boundary conditions.
I assume that the elastica does not stretch and preserves 
its local infinitesimal length.

I will consider quantization of 
a closed elastica whose local length preserves in the 
quantization process.
The partition function of the elastica is given as [25],
$$
      \Cal Z=\int D X \exp
       \left(-\beta \int^L_0 \dd s\left[ k^2\right] \right)  .
           \tag 2-5
$$
Since there is trivial affine symmetry of the centroid of 
the elastica and the partition function diverges, 
I will regularize it,
$$
            \Cal Z_{\reg}=\frac{\Cal Z}{\text{Vol}(\text{Aff})},
             \tag 2-6
$$
where $\text{Vol}(\text{Aff})$ is the volume of the affine 
transformation.

Next I will consider the condition of local length preserving.
In the path integral, I must pay attentions upon the 
higher perturbations of $\epsilon$ 
to gain an exact result.
Hence I will assume that $X$ is parameterized by a parameter $t$
and the difference between perturbed affine vector $X_\epsilon$
 and unperturbed one $X$ 
can be expressed by [6,25-27],
$$
       X_\epsilon(s,t):=\ee^{\epsilon \partial_{t}}X_{\qcl}(s,t),
       \quad \epsilon \partial_{t} X  =X_\epsilon - X 
       + \Cal O(\epsilon^2). \tag 2-7
$$
with the relation
$$
   \partial_{t} X_{\qcl} = (u_1 + \ii u_2 )\exp(\ii \phi_{\qcl}) ,
   \quad u_1(L)=u_1(0), \quad u_2(L)=u_2(0),
     \tag 2-8
$$
where $u$'s are real function of $s$ and $t$.
This is virtual dynamics of the curve [6].
As well as the argument in refs.[6,25-27], due to the isometry 
condition, I require $ [\partial_{t},\partial_s]=0$ for $X$. 
Then the isometry condition exactly
preserves, $\dd s\equiv \dd s_\epsilon$ for 
$\dd s_\epsilon :=\sqrt{ \partial_s 
\bar X_\epsilon\partial_s  X_\epsilon}\dd s$.  
Even though $\epsilon$ is constant, 
dependence of the variation upon the position $s$ comes from 
the "equation of motion" (2-8) and $u_a(s)$, $a=1,2$. 
Hence the deformation (2-7) contains non-trivial ones.

From $ [\partial_{t},\partial_s]=0$, I have  the relation [26,27],
$$
   - \partial_{t} \exp(\ii \phi_{\qcl}) 
   = \left((u_{1s}-u_2 k_{\qcl})
    +\ii(u_{2s}+u_1 k) \right)  \exp(\ii \phi_{\qcl})     .
                     \tag 2-9
$$
Noting that $\phi$ and $u$'s are real valued,
(3-9) is reduced to two differential equations 
and by partially solving one of them,
I obtain the "equation of motion" of the deformation, 
$$
           \partial_s u_1 = k_{\qcl} u_2, 
      \quad u_1 =\int^s \dd s \ u_2 k_{\qcl} 
      =: \partial_s^{-1} u_2 k_{\qcl} ,
$$
$$
           \partial_t k = \Omega u_2.
        \tag 2-10
$$
Here $\partial^{-1}_s$ is the pseudo-differential operator 
with a parameter $c\in\Bbb R$ as an integral constant  and
$$
 \Omega:=  \partial_s^2 +\partial_s 
 k_{\qcl}\partial_s^{-1} k_{\qcl} . \tag 2-11
$$
is the Gel'fand-Dikii operator for the MKdV equation [26,27].

In ref.[6,25], instead of the single deformation parameter, 
I used the infinite dimensional parameters 
$\bold t=(t_1,t_3,\cdots)$ and investigated the moduli space 
of the partition function.
Then the minimal set of the virtual equations of motion, 
which satisfies the physical requirements, is given as
$$
       \partial_{t_{2n+1}} k_{\qcl} 
       =- \Omega^n \partial_s k_{\qcl} , \quad 
      \partial_{t_{2n+3}} k_{\qcl}
      =\Omega \partial_{t_{2n+1}} k_{\qcl}, \quad
           (n=1,2,\cdots).
           \tag 2-12
$$
They are  the MKdV hierarchy [27,28]. As in ref.[6], 
I stated that these relations (2-12)
should be regarded as the Neither currents for the 
immersed object and
$t$'s should be considered as the Schwinger proper times, 
in ref.[25] I showed that (2-12) means 
the quantum fluctuations and a kind of currents 
of the quantized Neither
theorem or the Ward-Takahashi identities.

However by the studying the moduli of the quantize elastica,
the nontrivial deformation obeys the MKdV equation
$$
     \partial_t v + 6 v^2 \partial_s v + \partial_s^3 v=0,
     \tag 2-13
$$
because the solutions of the  higher order equations 
belonging to the MKdV hierarchy  
are also satisfied with  the MKdV equation.

Here it is a very remarkable fact that for the variation 
of $t$ obeying the MKdV equation,
the Bernoulli-Euler functional is invariant,
$$
  \partial_t  \int \dd s\ v(s,t)^2
  =\frac{1}{4}\partial_t E =0, \tag 2-14
$$
because
$$
 \partial_t\int \dd s v^2=- \int \dd s \partial_s 
 (\frac{3}{2} v^4
 +\frac{1}{2}(v \partial_s^2 v-(\partial_s v)^2)=0. \tag 2-15
$$

Since the MKdV problem is an initial value problem, 
for any regular shape of elastica
satisfied with the boundary conditions, the "time" $t$ 
development of the curvature can be expressed. 
In other words, for given any regular curve,
there exists family of the solutions of the MKdV equation
(2-13) which contains the given curve. 
Due to the integrability and (2-15),
during the motion of $t$, the energy functional does not 
change its value.
Hence the trajectory of the deformation parameter $t$ draws the 
functional space which has the same of value of the 
energy functional.
This remind me of the fact that in the group theory 
the character of a group is invariant among the elements belonging 
the same conjugate class. In fact in the Sato theory, the solutions
space of the MKdV equation is acted by the affine Lie algebra
 $A^{(1)}$ [31].

Thus I can estimate the functional space for each functional value.
In other words by investigating the moduli of the MKdV equation 
which is satisfied with
the boundary conditions,
$$
       v(0)=v(L), \quad      X(0)=X(L), \tag 2-16
$$
the measure of the functional integral $\dd \mu$ can be decomposed,
$$
            \dd \mu= \sum_E \dd \mu_E. \tag 2-17
$$
So I let the set of these trajectories which occupy 
the same energy $E$ be denoted as $\Xi_E$.

Hence the partition function can be represented as
$$
           \Cal Z_\reg = \int \dd \mu \exp(-\beta E )
           =\sum_E \exp(-\beta E) \int_{\Xi_E} 
           \dd \mu_E =\sum_E \exp(-\beta E) \text{Vol}(\Xi_E)
                      \tag 2-18
$$
where 
$$
           \text{Vol}(\Xi_E)= \int_{\Xi_E} \dd \mu_E \tag 2-19
$$
is the volume of the trajectories $\Xi_E$.

In ref.[25], I explicitly expressed $\dd \mu$ in terms of 
the moduli of the MKdV equation. According to the arguments 
in ref.[25], for a case of the solution represented by 
the hyperelliptic function of
genus $g$, $\dd \mu_E$ is roughly expressed as
 $\dd t_3 \wedge \dd t_5 \wedge$
$\cdots \wedge \dd t_{2g-1}$ where 
$\bold t_g:=(t_1,t_3,\cdots,t_{2g-1})$
is a subset of the infinite dimensional deformed parameters 
such as (2-12).
Even though I introduced the infinite dimensional coordinates
 $\bold t$ in ref.[25], they are often reduced to finite
  dimensional space, as the Jacobi variety with
finite dimension is embedded in the universal grassmannian manifold 
in the Sato theory [25,31,32]. I showed that 
 $\Xi_E$ is given as the 
real subspace of the Jacobi variety corresponding to the  
hyperelliptic curve, which is the trajectory space of the 
solution [25].

I will note that the volume of $\Xi_E$ is estimated by the unit of 
the elastica  length $L$. Due to the 
complex structure of the moduli of the MKdV equation, 
which is expressed as the 
Jacobi variety of the hyperelliptic curve and can be performed 
the coordinate transformation such as rotation,  the volume
can be evaluated in terms of the elastica length $L$ [25,32].

However since the dimension of the trajectory space $\Xi_E$ 
differs depending upon 
the energy $E$,  the sum of terms with  different dimensional 
volume appear. It seems to be fancy
but noting the facts that the dimension of the energy functional 
$E$ is the inverse of the length and 
that $\beta /$[length] is order unit, the multiple of the length
 can be interpreted as
the multiple of the quantizing parameter $\beta^{-1}$.
Hence such summation has physical meanings.

(2-18) means that  the density of state of the Bernoulli-Euler 
functional system is completely represented by the moduli and 
solutions spaces of the MKdV equation.
These space is acted by the infinite dimensional Lie group [31].
Then the measure $\dd \mu$ can be regarded as the Haar measure for 
the subalgebra of 
infinite dimensional Lie algebra $A^{(1)}$ [25].

\tvskip
\centerline{\twobf \S 3. Quantization of  Willmore surface}
\tvskip

 I will denote by $\SS$ a shape of a compact surface immersed
in the three dimensional space 
$\Bbb R^3 \approx \Bbb C \times \Bbb R$,
and by $(Z(z,\bar z):=X^1+\ii X^2,X^3(z,\bar z))$
 its affine vector : 
$$
   \Sigma \ni z \mapsto (Z,X^3) \in \SS \subset 
   \Bbb C \times \Bbb R.        \tag 3-1
$$
Here $\Sigma$ can be expressed as $\Sigma = \Bbb C/\Gamma$
where $\Gamma$ is a Fuchian group and then $\Sigma$ is a 
complex analytic object [33].
The volume element and the infinitesimal length the surface
 $\SS$ are given by 
$$
     \dd \text{vol}=\frac{\ii}{2}\rho\dd z\wedge \dd \bar z
     =:\frac{\ii}{2}\rho \dd^2 z, \quad
           \dd s^2= \rho \dd z \dd \bar z.
           \tag 3-2
$$

The Konopelchenko-Kenmotsu-Weierstrass-Enneper (KKWE) relation 
will be expressed as
$$
           \psi_+=\ii \sqrt{\bar \partial_z\bar Z}, 
           \quad\psi_-=-\ii \sqrt{ \partial_z\bar Z},
    \quad \partial X^3 = - \bar \psi_+ \bar \psi_- ,\tag 3-3
$$
$$
           \DD \psi =\pmatrix \partial & -V \\
        V & \bar \partial \endpmatrix \pmatrix \psi_+\\ 
        \psi_-\endpmatrix =0 , \quad 
                    V=\frac{1}{2}H \rho,
                    \tag 3-4
$$
where $V$ is a real valued function of $z$ and $\bar z$ 
and $H$ is the mean curvature of $\SS$. 
The general form might be expressed by the Dirac operator
which was calculated by Burgess and Jensen following
my prescription of the confinement Dirac operator. 
However due to the conformal structure of the surface $\SS$, 
the equation becomes simpler and given by (3-4) [16]
while due to the isometry condition, the Frenet-Serret relation
 becomes simple (2-2).

As the energy integral of the Willmore surface or the Polyakov
 extrinsic curvature action is given as
$$
           E=\int \rho \dd^2 z \ H^2=4\int \dd^2 z\  V^2 . 
           \tag 3-5
$$
The Willmore surface is realized as its stationary point.

As I quantized the elastica using the MKdV equation [25],
I will consider quantization of such a surface.
The partition function of the surface  is also given as [25],
$$
           \tilde {\Cal Z}_\reg =\frac{\int D X \exp\left(-\beta
             \int \rho \dd^2 z H^2\right)}
        {\text{Vol}(\text{Aff})}  ,
           \tag 3-6
$$
where $\text{Vol}(\text{Aff})$ means the volume of the affine
 transformation in $\Bbb R^3$.

I will search for the deformation flow of the surface which
 preserves the 
Willmore function or the Polyakov extrinsic curvature action and
complex structure. My question is what equation the deformation
flow obeys.
By Taimanov and Konopelchenko, the answer of  the question was
founded that the modified Novikov-Veselov (MNV) equation preserves
the complex structure and the functional (3-5) [10-14].

The MNV equation is given as
$$
           V_t=V_\tp+V_\tm,
$$ $$
   V_\tp=\partial^3 V+3 \partial V U+\frac{3}{2}V \partial U, \quad
  V_\tm=\bar \partial^3 V+3 \bar \partial 
  V \bar U+\frac{3}{2}V \bar \partial \bar U,
$$ $$
 \bar \partial U=\partial V^2 ,\quad  
 \partial \bar U=\bar\partial V^2.
           \tag 3-7
$$
Along the line $z = \bar z$, the MNV equation (3-7) 
is reduced to the MKdV equation (2-13).

As the Frenet-Serret relation can be regarded as the inverse 
scattering system of the MKdV equation, 
the KKWE equation can be also regarded as the inverse scattering 
system of the MNV equation.
$$
           (\partial_\tpm -B^\pm)\DD +[\DD,A^\pm]=0, \tag 3-8
$$
recovers (3-7) for
$$
    A^+= \pmatrix \partial^3 & -3(\partial V) \partial +3 VU\\
    0 & \partial^3 +3U \partial  +3(\partial U)/2 \endpmatrix,
$$
$$
           B^+=3\pmatrix 0 & (\partial V)\partial - VU \\
          -(\partial V)\partial-(\partial^2 V)-UV & 0 \endpmatrix,
$$
$$
  A^-= \pmatrix \bar \partial^3 + \bar U \bar \partial 
  +3\bar \partial \bar U/2 & 0\\
   3\bar \partial V \bar \partial -3V \bar U & \bar \partial^3 
    \endpmatrix,
$$
$$
           B^-=3\pmatrix 0 & (\bar \partial V)\bar
            \partial +(\bar \partial^2 V)-V\bar U  \\
        -(\bar \partial V)\bar \partial+ V\bar U & 0 \endpmatrix.
         \tag 3-9
$$
The variation of the Dirac field is given as
$$
   \partial_t\psi=\partial_\tp\psi+  \partial_\tm\psi,\quad
   \partial_\tpm\psi  = A^\pm \psi. \tag 3-10
$$

For the variation of $t$ obeying the MNV equation, the Willmore 
functional is invariant,
$$
    4 \partial_t  \int \dd^2 z\  V^2=\partial_t E =0. \tag 3-11
$$
because the integrand can be expressed by the
 boundary quantities [13],
$$
      V^2_t = \partial (V \partial^2 V-\frac{1}{2} 
      (\partial V)^2 +\frac{3}{2}V^2 U)+
      \bar \partial (V \bar\partial^2 V-\frac{1}{2}
       (\bar\partial V)^2 +\frac{3}{2}V^2\bar U). \tag 3-12
$$

Next I will check the preserving the complex structure of 
surface for the MNV flows
following the argument of Taimanov.

First I will remark that the metric is represented by the
 Dirac field as
$$
           \rho = (|\psi_1|^2+|\psi_2|^2)^2, \tag 3-13
$$
owing to the relation (3-3). Thus if the relation (3-3) is covariant 
or preserves for a point of the MNV flows, 
the conformal structure (3-13) maintains.

Thus I will evaluate $\partial_t Z=\partial_\tp Z+\partial_\tm Z$
 and $\partial_t X^3$.
By straight forward computations, these values calculated as [13]
$$
           \partial_\tpm Z =2 \ii  \int^{z(\bar z) } 
           \dd (f_\pm+g_\pm), \tag 3-14
$$
and 
$$
           \partial_t X^3 =- \int^{z(\bar z) } 
           \dd (h_1+h_2) \tag 3-15
$$
where
$$
           \split
           f_+ &:=\frac{3}{2} U \psi_-^2, \quad 
           g_+ := \psi_- \partial^2 \psi_-
                      -\frac{1}{2} (\partial \psi_-)^2\\
           f_- &:=\frac{3}{2} \bar U \psi_+^2, \quad 
           g_- := \psi_+ \bar \partial^2 \psi_+
                      -\frac{1}{2} (\bar \partial \psi_+)^2
                      ,\endsplit
                       \tag 3-16
$$
$$
      \split
      h_1&= \bar \psi_+ \partial^2 \psi_- + \psi_- 
     \partial^2 \bar \psi_+ -\partial \psi_-\partial
      \bar \psi_+ + 3 U\bar \psi_+ \psi_- ,\\
      h_2&=  \psi_+ \bar\partial^2 \bar\psi_- + 
      \bar\psi_- \bar\partial^2 \psi_+ - 
           \bar\partial  \bar\psi_- \bar\partial  \psi_+ + 
           3 \bar U \psi_+\bar \psi_-  .\endsplit \tag 3-17
$$
%$\partial_+:=\partial$, $\partial_-:=\bar \partial$,
 $U^{(+)}=U$ and $U^{(-)}=\bar U$.
Here $\dd f= \partial f \dd z+ \bar \partial f \dd \bar z$.

Let the infinitesimal flows obeying the MNV equation (3-7) 
module $\epsilon^2$ be denoted as
$$
(Z_\epsilon,X_\epsilon^3):=(Z,X^3)+
\epsilon \partial_t(Z,X^3)+ \Cal O(\epsilon^2). \tag 3-18
$$
(3-14)-(3-17) means the infinitesimal variation is 
given as the integral of the closed form defined over 
$\Sigma$ [34]
 and can be regarded as a single function of $\Sigma$. 
Since $(Z,X^3)$ is also a periodic function of $\Sigma$, 
$(Z_\epsilon,X_\epsilon^3)$ 
is globally defined over $\Sigma$ as a function of $\Sigma$.

On the other hand, (3-14)-(3-17) guarantee
that $[\partial,\bar \partial]X^i_\epsilon=0$, which
 means that I can locally define the independent 
coordinates $z$ and $\bar z$ for $X^i_\epsilon$ surface;
I can locally find a conformal coordinate system of
 an open set of $\Sigma$. Furthermore due to the global
  properties, their coordinate system can be extended to the
global coordinate and the connection of each open set are
 trivial due to $[\partial,\bar \partial]X^i_\epsilon=0$
  for any point of $\Sigma$.

Hence the MNV flows preserves the complex structure of the surface 
$\SS$.

I will emphasis that
 the MNV problem is also an initial value problem, for any shape of
  compact conformal surface, the "time" $t$ development of
   the surface can be expressed
  and these conserves the energy functional and complex structure.
Hence the trajectory of the deformation parameter $t$ 
means the states of the same energy and its volume is the 
density of states of each energy $E$.

As I did for the quantization elastica,
the measure of the functional integral $\dd \mu$ can be decomposed,
$$
     \dd \tilde \mu= \sum_E \dd \tilde \mu_E   \tag 3-19
$$
and moduli of $\tilde {\Cal Z}_\reg$ restricted by $E$ is 
denoted as $\tilde \Xi_E$
$$
    \tilde {\Cal Z}_\reg= \int \dd \tilde 
     \mu \exp(-\beta E )=\sum_E \exp(-\beta E) 
     \int_{\tilde \Xi_E} \dd\tilde  \mu_E
     =\sum_E \exp(-\beta E) \text{Vol}(\tilde \Xi_E) \tag 3-20
$$
As I did for the Bernoulli-Euler functional, the density of state 
of the Willmore functional system
might be completely represented by the moduli of the MNV equation.

However the Willmore surface has no natural length because for 
a global scale transformation $z \to \lambda z$ 
($\lambda >0$), the mean curvature changes as $H \to H/\lambda$ 
and the Willmore surface is invariant.
Hence for given energy $E$, there are infinite degenerate states 
regarding to the global scaling parameter $\lambda \in (0,\infty)$
 and the regularized partition function
$\tilde {\Cal Z}_\reg$ also diverges.

However along the line of $z=\bar z$, the deformation of MNV flows
 obeys the MKdV equation
which conserves local length of the line. In other words, 
on the MNV flows, 
the length of the line is a conserved quantity and is well-defined.
 Hence in terms of this length, I can redefine the partition 
 function by fixing the length of the line,
$$
   {\Cal Z}_\reg:=\tilde {\Cal Z}_\reg |_{(\text{the length of }z
   =\bar z) =L}. \tag 3-21
$$
Due to the compactness of the surface $\SS$, $L$ must be finite. 
By fixing the scale of the surface, I will define  decomposed 
measure and the space of the trajectories,  
$\dd \mu_{E,L}:=\dd \tilde \mu|_L$, $\Xi_{E,L}:={\tilde \Xi_E}|_L$,
$$
    {\Cal Z}_\reg =\sum_E  \int_{ \Xi_{E,L}} \dd\tilde
      \mu_{E,L}\exp(-\beta E)
    =\sum_E \exp(-\beta E) \text{Vol}( \Xi_{E,L}) \tag 3-22
$$
The physical meaning of the summation in (3-22) is justified
 similar to (2-19). 

\tvskip
\centerline{\twobf \S 4. Discussion}
\tvskip

Carroll and Konopelchenko also proved that the MNV flows conserves
 the extrinsic string action for the case that $\rho H$ 
 is constant; here the extrinsic string action consists of the
Nambu-Goto action, the Wess-Zumino-Witten type geometrical action,
 and the Polyakov extrinsic curvature action (3-3).
Hence my result (3-2)  can be extended to such a case and then
 it means that the algorithm of the
calculation of the partition function of the extrinsic string
 in $\Bbb R^3$ is essentially 
the same as above arguments.
In other words, the quantization of the string immersed
 in $\Bbb R^3$ can be  partially performed even though
only the string in $\Bbb R^n$ $n<3$ had been studied as 
the two dimensional gravity [28-30].

In ref. [2] and [3], I stated that as the self-dual Yang-Mills
 equation can be expressed by 
the integrable equation and be represented by the Dirac operator
 and as the MKdV equation governs the "virtual" motion of
  the elastica and can be 
written by the Dirac operator, the higher dimensional
soliton surface might be expressed by the Dirac operator
 and has a physical meaning.

This conjecture was proved by the discoveries and studies of
 Konopelchenko and Taimanov.
Using the Dirac operator, they investigated the surface itself
 and derived non-trivial results [10-14].
Categories of the linear analytic system of Dirac operator,
 the geometry and integrable system are closely connected with
  each other. 
Their connections should be interpreted as functors among them
and the exact sequences and the associated cohomorogy in individual
 categories should be expressed by common language.
Thus the expression of these systems should be unified and then
 using the relations their common hidden symmetries  might be
  revealed. I believe that this quantization of surface and
   quantization of elastica [25]
contributes such studies.

Finally I will comment upon open problems related to this system.
In ref.[25], I found that 
at the critical point of the quantized elastica, a certain
 expectation value obeys the Painlev\'e
  equation of the first kinds.
I have a question what equation appears at a critical point  
in the quantized Willmore surface system. If exists, it might
be related to the higher dimensional analogue of the Painlev\'e
 equation.

Furthermore since the MNV equation is an initial value problem,
 more general Riemannian surfaces can be allowed, at least,
  an initial condition even though the energy 
manifold of the inverse scattering system is given as only
 hyperelliptic curves [6].
Hence I have another question  whether there  is an analytical
 connection between the general Riemannian surface or
  the general Fuchian group and the hyperelliptic function
   of the MKdV equation.
If there is, this system should be algebraically studied.

%\newpage

\tvskip
\centerline{\twobf Acknowledgment}
\tvskip

I would like to thank  Y. \^Onishi
for helpful discussions and continuous encouragement
and Prof. B.~G.~Konopelchenko for sending me his interesting works.
I also acknowledge that the seminars on
differential geometry, topology, knot theory and group theory with
Prof. K.~Tamano influenced this work.

%\enddocument
%\newpage

\enddocument